\documentclass[aps,prb,twocolumn,nofootinbib,floatfix]{revtex4}
\usepackage{amsmath}
\usepackage{epsfig}
\newcommand{\beq}{\begin{equation}}
\newcommand{\eeq}{\end{equation}}
\newcommand{\bqn}{\begin{eqnarray}}
\newcommand{\eqn}{\end{eqnarray}}
\newcommand{\bqns}{\begin{eqnarray*}}
\newcommand{\eqns}{\end{eqnarray*}}
\newcommand{\bary}{\begin{array}}
\newcommand{\eary}{\end{array}}

\begin{document}

\title{Finger-gate array quantum pumps: pumping characteristics and mechanisms}
\author{S. W. Chung$^{1}$, C. S. Tang$^{2}$, C. S. Chu$^{3}$, and C. Y. Chang$%
^{1}$}
\affiliation{$^{1}$Department of Electronics, National Chiao-Tung University, Hsinchu
30010, Taiwan\\
$^{2}$Physics Division, National Center for Theoretical Sciences,
P.O. Box 2-131, Hsinchu 30013, Taiwan\\
$^{3}$Department of Electrophysics, National Chiao-Tung
University, Hsinchu 30010, Taiwan}
\date{\today }

\begin{abstract}
\ \ We study the pumping effects, in both the adiabatic and
nonadiabatic regimes, of a pair of {\it finite} finger-gate array
(FGA) on a narrow channel. Connection between the pumping
characteristics and associated mechanisms is established. The
pumping potential is generated by ac biasing the FGA pair. For a
single pair ($N=1$) of finger gates (FG's), the pumping mechanism
is due to the coherent inelastic scattering of the traversing
electron to its subband threshold. For a pair of FGA with pair
number $N>2$, the dominant pumping mechanism becomes that of the
time-dependent Bragg reflection. The contribution of the
time-dependent Bragg reflection to the pumping is enabled by
breaking the symmetry in the electron transmission when the
pumping potential is of a predominant propagating type. This
propagating wave condition can be achieved both by an appropriate
choice of the FGA pair configuration and by the monitoring of a
phase difference $\phi$ between the ac biases in the FGA pair. The
robustness of such a pumping mechanism is demonstrated by
considering a FGA pair with only pair number $N=4$.
\end{abstract}

\maketitle
%%%%%%%%%%%%%%%%%%%%%%%%%%%%%%%%%%%%%%%%%%%%%%%%%%%%%%%%%%

\section{INTRODUCTION}

Quantum charge pumping (QCP) has become an active field in recent
years.~\cite{Thou83,Niu90,Hekking91,Brouwer98,Swit99,Alei98,Zhou99,Wohl02,
Wei00,Tang01,Poli01,Avro01,Wang02,Mosk01,Mosk02,Kim03,Mosk03H,Mosk03,
Cohen03,Shar03} This is concerned with the generation of net
transport of charges across an unbiased mesoscopic structure by
cyclic deformation of two structure parameters. Original proposal
of QCP, in the adiabatic regime, was due to
Thouless.~\cite{Thou83,Niu90} He considered the current generated
by a slowly varying travelling wave in an isolated one-dimensional
system.  The number of electrons transported per period was found
to be quantized if the Fermi energy lies in a gap of the spectrum
of the instantaneous Hamiltonian.  Aiming at this quantized pumped
charge nature of the adiabatic pumping, Niu proposed various
one-dimensional periodic potentials for the adiabatic quantum
pumping (AQP),~\cite{Niu90} and pointed out the importance of the
quantized charge pumping in utilizing it for a direct-current
standard.~\cite{Niu90}

Another way to achieve the AQP was suggested by Hekking and
Nazarov,~\cite{Hekking91} who studied the role of inelastic
scattering in the quantum pumping of a double-oscillating barrier
in a one-dimensional system. Intended to stay in the adiabatic
regime, they invoked a semiclassical approximation and had assumed
that the Fermi energy $\varepsilon_{\rm F} \gg \hbar\Omega$, where
$\Omega$ is the pumping frequency. This semiclassical treatment of
the inelastic scattering is known to be inappropriate for the
regime when either the initial or the final states are in the
vicinity of the energy band edge. Such a regime, however, is our
major focus in this work.  It is because the coherent inelastic
scattering becomes resonant when the traversing electron can make
transitions to its subband threshold by emitting
$m\hbar\Omega$.~\cite{Bag92,Reichl01}  Depend on the system
configuration, this and another resonant inelastic scatterings
will be shown to dominate the pumping
characteristics.~\cite{Tang01}

A recent experimental confirmation of AQP has been reported by
Switkes \textit{et al}.~\cite{Swit99} Two metal gates that defined
the shape of an open quantum dot were ac biased~\cite{Tang03} with
voltages of the same frequency but differed by a tunable phase
difference.~\cite{Brouwer98} DC response across the source and
drain electrodes is the signature of the AQP. This has prompted
further intensive studies on AQP in: quantum
dots,~\cite{Alei98,Zhou99,Wohl02} double-barrier quantum
wells,~\cite{Wei00} pumped voltage,~\cite{Poli01} noiseless
AQP,~\cite{Avro01} heat current,~\cite{Wang02} incoherent
processes,~\cite{Mosk01,Mosk02} quantum
rings,~\cite{Mosk03,Cohen03} and interacting wires.~\cite{Shar03}

An alternate experimental effort in generating AQP involves
surface acoustic wave
(SAW).~\cite{Shi96,Tal97,Wohl00,Wohl01,Wohl02R} Generated by an
interdigitated SAW transducer located deep on an end-region of a
narrow channel, the SAW propagates to the other end-region of the
narrow channel while inducing a wave of electrostatic potential
inside the channel. Electrons trapped in the potential minima are
thus transported along the narrow channel.  Both Mott-Hubbard
electron-electron repulsion in each such trap and the adiabaticity
in the transport are needed to give rise to quantization in the
pumped current.~\cite{Tal97} As such, the channel has to be
operated in the pinch-off regime.~\cite{Wohl00}

\begin{figure}[t]
\includegraphics[width=.3 \textwidth]{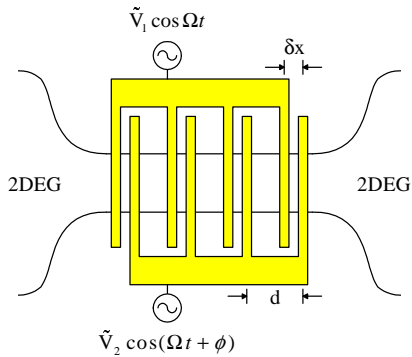}
\caption{(color online). Top view of the proposed system structure
for the case of pair number $N=4$. A FGA pair locates on top of a
narrow channel. $\tilde{V}_{i}$ denotes the amplitude of the
potential energy, and $\phi$ is the phase difference.}
\label{Fig:sys}
\end{figure}

In this work, we propose to study yet another experimental
configuration for QCP in a narrow channel. The proposed
configuration consists of a pair of {\it finite\/} FGA, with the
number $N$ of FG's in each FGA being kept to a small number. In
contrast to the SAW configuration, the FGA pair sits on top of the
narrow channel, rather than locating at a distance far away from
it; and the most significant QCP occurs in regimes other than the
pinch-off regime. The FG's orient transversely and line up
longitudinally with respect to the narrow constriction. As is
shown in Fig.~\ref{Fig:sys}, pumping potential can be generated by
ac biasing the FGA pairs with the same frequency but maintaining a
phase difference $\phi$ between them. Since the wave of
electrostatic potential induced in the narrow channel is directly
from the FG's, rather than via the SAW, our proposed structure has
the obvious advantage that the working frequency is not restricted
to the frequency of the SAW, $\omega_{\rm S}=2\pi v_{\rm S}/d$.
Here $v_{\rm S}$ is the phase velocity of the SAW, and $d$ is the
pitch in the FGA. Furthermore, when the working frequency is
different from $\omega_{\rm S}$, the contribution from SAW to the
pumped current will be negligible.

Below we shall show how the ac biased FGA pair plays a subtle role
in the generation of QCP. In Sec. II, we present our theoretical
model for the FGA pair calculation of the pumped current generated
by the FGA pair configuration. In Sec. III, we present the pumping
characteristics and demonstrate that resonant coherent inelastic
scatterings are the underlying pumping mechanisms.
 Finally, in Sec. IV, we present our discussion and
summary.

\section{FGA pair model}

The potential $V(x,t)$ in a narrow constriction induced by a FGA
pair is represented by \beq V(x,t)=\sum\limits_{i=1}^{N}\
V_{1i}(x)\cos (\Omega t)
+ V_{2i}(x)\cos (\Omega t + \phi),%
\eeq%
where $N$ is the number of FG's per FGA. We assume that the ac
biased FGA pair are localized, respectively, at positions $x_i$
and $x_i + \delta x_i$, namely that $V_{1i}(x)= V_{1}\delta
(x-x_{i})$ and $V_{2i}(x)=V_{2}\delta (x-x_{i}-\delta x)$ with a
relative phase difference $\phi$. These FG's are evenly spaced,
with a pitch $d$, and are located at $x_{i}=(i-1)d$ for one FGA
and $x_{i}+\delta x$ for the other. The relative shift between the
FGA pair is $\delta x=\alpha \, d$, where the fractional shift
$0<\alpha <1$. In the following, we consider the case of the same
modulation amplitude $V_{1}=V_{2}=V_{0}$. Depending on the choice
of the values for $\phi $ and ${\alpha}$, $V(x,t)$ will either be
predominantly of a propagating or a standing wave type. A sensible
choice can be made from considering the lowest order Fourier
component of $V(x,t)$, given by
\beq {\cal V}_1 = \frac{2V_{0}}{d}
\left\{ \cos K x \cos \Omega t + \cos \left[ K(x-\delta x) \right]
\cos (\Omega t+\phi) \right\}, \eeq
 where $K= 2\pi/d$. For our purposes in
this work, an optimal choice is $\phi =\pi/2$ and $\alpha =1/4$,
in which $V(x,t)$ is a predominant left-going wave.

The Hamiltonian of the system is $H=H_{y}+H_{x}(t)$, in which
$H_{y}=-\partial ^{2}/ \partial y^{2} + \omega _{y}^{2}y^{2}$
contains a transverse confinement, leading to subband energies
$\varepsilon _{n}=(2n+1)\omega _{y}$. The time-dependent part of
the Hamiltonian $H_{x}(t)$ is of the dimensionless form $H_{x}(t)$
= $-\partial ^{2}/\partial x^{2} + V(x,t)$. Here appropriate units
have been used such that all physical quantities presented are in
dimensionless form.~\cite{Tang03}

In the QCP regime, the chemical potential $\mu $ is the same in
all reservoirs.  Thus the pumped current, at zero temperature, can
be expressed as~\cite{Tang01} \beq
I=-\frac{2e}{h}\int_{0}^{\mu}dE\,\left[T_{\rightarrow
}(E)-T_{\leftarrow}(E)\right].
\eeq%
Here the total current transmission coefficients include the
contributions by electrons with incident energy $E$ in incident
subband $n$, which may absorb or emit $m\Omega$ to energy $E_m =
E+m\Omega$ by the FG pumping potentials, given by \beq
T_{\rightarrow (\leftarrow )}(E)=\sum_{n=0}^{{\cal N}_{\rm S} -
1}\sum_{m=-\infty}^{\infty} T_{n\rightarrow (\leftarrow )}(E_m,E),
\eeq where ${\cal N}_{\rm S}$ stands for the number of occupied
subbands. The summations are over all the propagating components
of the transmitted electrons, and includes both the subband index
$n$ and the sideband index $m$. The subscripted arrow in the total
current transmission coefficient indicates the incident direction.
These coefficients are calculated numerically by a time-dependent
scattering-matrix method.~\cite{Tang01,Wag95,Tang00}

\section{NUMERICAL RESULTS}

In this section we present the numerical results for the pumping
characteristics of either a single FG pair $(N=1)$ or a
\textit{finite} FGA pair $(N>2)$.  In these two cases that the
pumping characteristics are due to different resonant inelastic
scattering processes.  For definiteness, the unit scales in our
numerical results are taken from the GaAs-Al$_{\rm x}$Ga$_{\rm
1-x}$As based heterostructure. The values that we choose for our
configuration parameters are $\omega _{y}=0.007$, subband level
spacing $\Delta \varepsilon=2\omega_{y}$ $(\simeq 0.13\,${\rm
meV}$)$, $d=40$ $(\simeq 0.32$ $\mu${\rm m}$)$, and
$V_{0}=0.04\,(\simeq 28.7\,{\rm meV}\,{\rm \AA})$. From the value
of $V_{0}$, and the assumed FG width $\sim 0.05\,\mu$m, the
amplitude of the potential induced by a FG is $\sim 0.057\,$mV.

\subsection{single FG pair case}

\begin{figure}[th]
\includegraphics[width=.4 \textwidth]{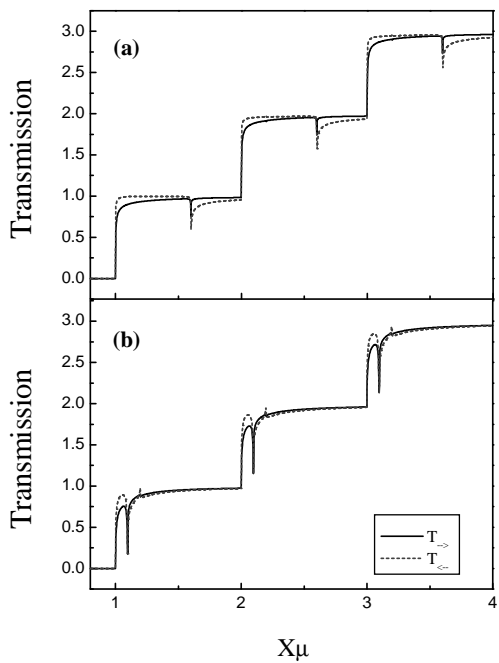}
\caption{Total current transmission coefficient versus X$_{\mu}$
for a pair of FG; (a) $\Omega =0.6\bigtriangleup \protect%
\varepsilon$ and (b) $\Omega =0.1\bigtriangleup \protect%
\varepsilon$. The transmission of the right-going (left-going)
electrons are represented by the solid (dotted) curve. The subband
level spacing is $\Delta \varepsilon$. Parameters $\alpha=1/4$ and
$\phi=\pi/2$ are chosen to meet the optimal condition.}
\label{Fig:N1T}
\end{figure}

In this subsection we investigate the pumping characteristics for
the case of a single FG pair. Figure \ref{Fig:N1T} presents the
dependence of the total current transmission coefficients on the
incident electron energy $\mu$. We replace the chemical potential
$\mu$ by \beq X_{\mu } = \frac{\mu}{\Delta \varepsilon}
+\frac{1}{2}\, ,
\eeq
which integral value corresponds to the
number of propagating subbands ${\cal N}_{\rm S}$ in the narrow
channel. The pumping frequency is higher in Fig.~\ref{Fig:N1T}(a),
with $\Omega=0.6\Delta\varepsilon$ $(\Omega/2\pi \simeq 18\,\,{\rm
GHz})$, than that in Fig.~\ref{Fig:N1T}(b), where
$\Omega=0.1\Delta\varepsilon$ $(\Omega/2\pi \simeq 3\,\,{\rm
GHz})$. We select the phase shift $\phi=\pi/2$ and $\alpha =1/4$.

At integral values of $X_{\mu}$, the total current transmission
coefficients $T_{\rightarrow}\,_{(\leftarrow)}(X_{\mu})$ exhibit
abrupt changes. This is due to the changes in the number of
propagating subbands in the narrow channel. Between integral
$X_{\mu}$ values, $T_{\rightarrow}\,_{(\leftarrow)}$ both show dip
structures. The dip structures are located at $X_{\rm dip}={\cal
N}_{\rm S}+0.6$ in Fig.~\ref{Fig:N1T}(a), and at $X_{\rm
dip}={\cal N}_{\rm S}+0.1$ in Figs.~\ref{Fig:N1T}(b). These dip
structure locations are the same for both $T_{\rightarrow}$ and
$T_{\leftarrow}$, and are resonant structures associated with
inelastic scattering that causes an electron to jump into a
quasibound state (QBS) just beneath a subband bottom.\cite{Bag92}
The peak structures in $T_{\leftarrow}$ of Fig.~\ref{Fig:N1T}(b),
and at $X_{\mu}={\cal N}_{\rm S}+0.2$, are $2\Omega$ resonant
structures.

\begin{figure}[th]
\includegraphics[width=.42 \textwidth]{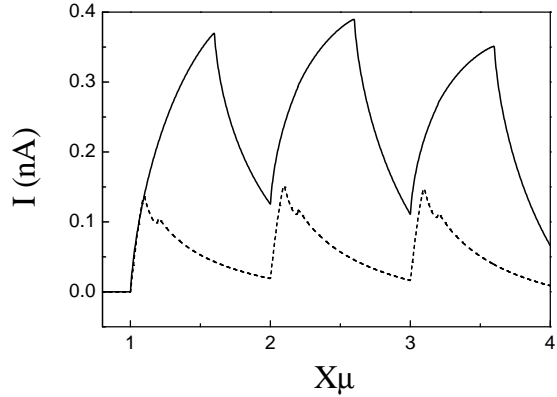}
\caption{The pumped currents versus $X_{\mu}$ with the same
parameters used in Fig.~\ref{Fig:N1T}. The solid, and dashed,
curves correspond, respectively, to $\Omega =0.6\Delta \varepsilon
$ and $\Omega =0.1\Delta\varepsilon $.} \label{Fig:N1C}
\end{figure}

In Fig.~\ref{Fig:N1T}, we can see that $T_{\leftarrow}(X_{\mu})$
does not equal to $T_{\rightarrow}(X_{\mu})$, this allows the
occurrence of the pumped current. Moreover, between integral
$X_{\mu}$ values, $T_{\leftarrow} > T_{\rightarrow}$ on the left
region of a dip structure, while $T_{\leftarrow}<T_{\rightarrow}$
on the right region of the dip structure. This has an important
bearing on the dependence of the pumped current on $\mu$, as is
shown in Fig.~\ref{Fig:N1C}. The pumped current rises, and drops,
on the left, and right, region of a $X_{\rm dip}$, respectively,
in accordance with the relative changes in $T_{\rightarrow}$ and
$T_{\leftarrow}$ about the same $X_{\rm dip}$. Hence the peaks of
the pumped current depend on the pumping frequency, at \beq
X_{\mu}^{\rm (peak)}={\cal N}_{\rm
S}+\frac{\Omega}{\Delta\varepsilon}\, , \eeq reassuring us that
the pumping is dominated by the aforementioned resonant inelastic
process.

Besides the trend that the pumped current in Fig.~\ref{Fig:N1C}
drops with the pumping frequency, we would like to remark on a
more interesting result: that both the adiabatic and nonadiabatic
behaviors can be found in the same curve. Since the adiabatic
condition is given by $\mu\gg \Omega$, the curve for
$\Omega=0.1\Delta\varepsilon$ in the regions ${\cal
N}_{s}+\Omega/\Delta\varepsilon<X_{\mu}\le {\cal N}_{s}+1$
corresponds to the adiabatic regimes, while the other $X_{\mu}$
regions are nonadiabatic regimes. This is checked also with our
other calculation, which is not shown here, using the Brouwer
expression.~\cite{Brouwer98} For the higher pumping frequency,
$\Omega=0.6\Delta\varepsilon$, the adiabatic condition is not
satisfied in the entire $X_{\mu}$ region, even though the pumping
characteristics resemble that of the adiabatic one in the regions
${\cal N}_{s}+\Omega/\Delta\varepsilon<X_{\mu}\le {\cal N}_{s}+1$.

\subsection{finite FGA case}

In this subsection we present the numerical results for the
pumping characteristics of a \textit{finite} FGA pair.   QCP for
two prominent modes of tuning the system are considered.  These
are (i) tuning of the electron density by the \textit{back-gate}
technique, and (ii) tuning of the channel width by
\textit{split-gate} technique.

\subsubsection{tuning back-gate}

\begin{figure}[th]
\includegraphics[width=.6 \textwidth,angle=270]{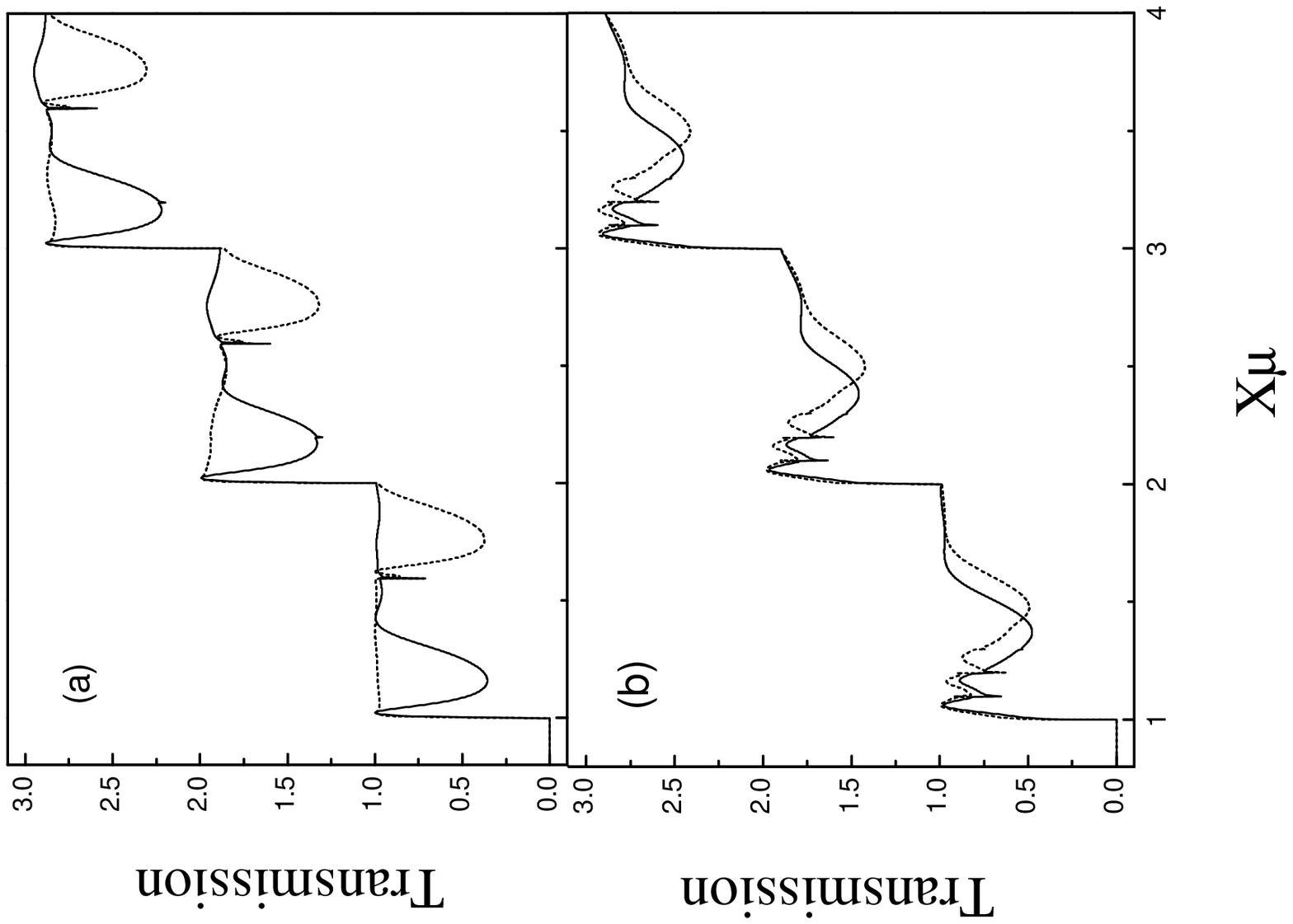}
\caption{Total current transmission coefficient versus X$_{\mu}$
for $N=4$; (a) $\Omega =0.6\Delta \varepsilon$ and (b) $\Omega
=0.1\Delta \varepsilon$. The transmission of the right-going
(left-going) electrons are represented by the solid (dotted)
curve.  The parameters $\alpha=1/4$ and $\phi=\pi/2$.}
\label{Fig:N4T}
\end{figure}

We present the numerical results for the pumping characteristics
of a FGA pair with $N=4$ that is realized by the \textit{back-gate
technique}. The dependence of the total current transmission
coefficients on $X_{\mu}$ is shown in Fig.~\ref{Fig:N4T}, in which
the pumping frequencies are (a) $\Omega=0.6\Delta\varepsilon$, and
(b) $\Omega=0.1\Delta\varepsilon$. The choice of the parameters
$d$, $\phi$, and $\alpha$ is the same as in the previous
subsection, but the latter two parameters give rise here to an
equivalent left-going wave in the pumping potential $V(x,t)$.

The curves in Fig.~\ref{Fig:N4T} show additional structures, other
than the dip structures that has been discussed in the last
subsection. These additional structures are valley structures that
occur at different $X_{\mu}$ values for $T_{\rightarrow}(X_{\mu})$
and $T_{\leftarrow}$. In a region between two integral values of
$X_{\mu}$, the valley structure of $T_{\rightarrow}(X_{\mu})$
occurs at a lower $X_{\mu}$. This shows clearly the breaking of
the transmission symmetry by the pumping potential. Furthermore,
the valleys are separated by $\Delta
X_{\mu}=\Omega/\Delta\varepsilon$.  This can be understood from
resonant coupling conditions
$\varepsilon_{k}=\varepsilon_{k-K}-\Omega$ and
$\varepsilon_{k+K}=\varepsilon_{k}-\Omega$ for, respectively, the
right-going and the left-going $k$. From these conditions, the
valley locations are at \beq k_{\pm}^{2} =
\left[\frac{K}{2}\left(1\mp
\frac{\Omega}{K^{2}}\right)\right]^{2}, \eeq
 where the upper sign
is for positive, or right-going, $k$. These locations, expressed
in terms of $X_{\mu}$, are given by \beq X_{\mu} = {\cal N}_{\rm
S} +
 \frac{k_{\pm}^{2}}{\Delta\varepsilon} , \eeq and are at $X_{\mu}$
= 1.19, 1.79, 2.19, 2.79, 3.19, and 3.79 for the case of
Fig.~\ref{Fig:N4T}(a), and $X_{\mu}$ = 1.39, 1.49, 2.39, 2.49,
3.39, and 3.49 for the case of Fig.~\ref{Fig:N4T}(b). The matching
between these numbers and our numerical results in
Fig.~\ref{Fig:N4T} is remarkable. In addition, energy gaps open up
at these $k_{\pm}^{2}$ locations, causing the drop in the
transmission and the formation of the valley
structures.~\cite{Tang01}  All these results reassure us that the
time-dependent Bragg's reflection is the dominant resonant
inelastic scattering in our FGA pair structure.

On the other hand, the adiabatic condition is here given by
$\varepsilon_{\rm gap}\gg\Omega$, where $\varepsilon_{\rm gap}$ is
the {\it effective \/} energy gap of the {\it instantaneous\/}
Hamiltonian.~\cite{Niu90} Since $\varepsilon_{\rm gap}$ is given
by the widths of the valley structures, therefore contributions of
the valleys to the pumped current is nonadiabatic in
Figs.~\ref{Fig:N4T}(a), because the valleys are well separated,
and adiabatic in Figs.~\ref{Fig:N4T}(b), because the valleys
overlap.

\begin{figure}[th]
\includegraphics[width=.3 \textwidth,angle=270]{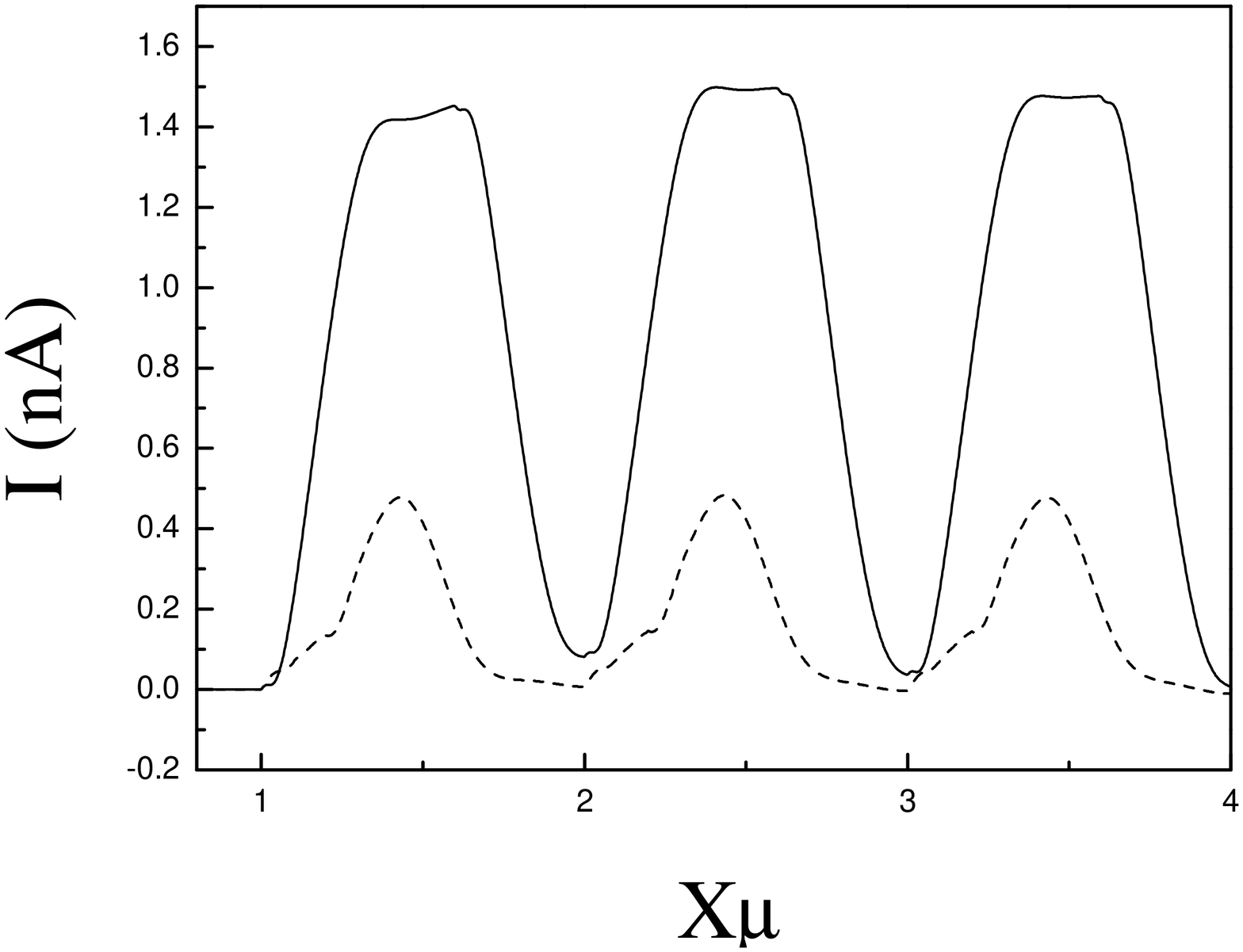}
\caption{Pumped current versus $X_{\mu }$. The choices of
parameters are the same as in Fig.~\ref{Fig:N4T}. The solid, and
dashed, curves correspond, respectively, to $\Omega =0.6\Delta
\varepsilon $ and $\Omega =0.1\Delta\varepsilon $.}
\label{Fig:N4C}
\end{figure}

In Fig.~\ref{Fig:N4C}, we present the $X_{\mu}$ dependence of the
pumped current for the cases in Fig.~\ref{Fig:N4T}. The pumped
current peaks at $X_{\mu}$ that lies in the middle between a
valley in $T_{\rightarrow}(X_{\mu})$ and the corresponding valley
in $T_{\leftarrow}(X_{\mu})$. The locations are around \beq
X_{\mu}={\cal N}_{\rm
S}+\frac{K^{2}}{4\Delta\varepsilon}\left(1\,+\,\frac{\Omega^{2}}
{K^{4}} \right), \eeq which depend on both the pitch $d$ and the
pumping frequency $\Omega$. The peaks have flat tops for the solid
curve, when $\Omega=0.6\Delta\varepsilon$. Comparing with the
total current transmission curves in Fig.~\ref{Fig:N4T}(a), we see
that the flat-topped peak profile is associated with the complete
separation between the valleys in $T_{\rightarrow}$ and
$T_{\leftarrow}$. This is in the nonadiabatic regime.  In
contrast, for the case when the valleys overlap, such as in
Fig.~\ref{Fig:N4T}(b), the pumped current peaks no longer carry a
flat-top profile, as is shown by the dashed curve in
Fig.~\ref{Fig:N4C}. This is in the adiabatic regime. Meanwhile,
their peak values are lowered. It is because cancellation sets in
when the valleys overlap. We note that the pumped currents are of
order nA.

The robustness of the time-dependent Bragg reflection, on the
other hand, is demonstrated most convincingly by the number of
charge pumped per cycle at the maximum  $I_{\rm Max}$ of the
pumped current.  In the dashed curve of Fig.~\ref{Fig:N4C}, the
pumped charge per cycle per spin state $Q_{\rm P} = (2\pi/\Omega)
I_{\rm Max}/2e = 0.495$, where $I_{\rm Max} = 0.48$ nA and $\Omega
= 0.1\Delta\varepsilon = 3.03$ GHz. To get a unity, or quantized,
charge pumped per cycle per spin state, one can fix the pumping
frequency $\Omega = 0.1 \Delta\varepsilon$, $N=4$, $\phi = \pi/2$,
and $d=40$, then tune the other pumping parameters as $V_0 = 0.09$
and $\alpha = 0.15$  to obtain $Q_{\rm P}=0.992$ at
$X_{\mu}=3.465$ (not shown here). In this frequency regime, the
pumping would be expected to be adiabatic, according to
Thouless~\cite{Thou83} and Niu~\cite{Niu90} when $\varepsilon_{\rm
gap}\gg\Omega$. However, in our case here, the energy gap is at
best only partially opened, as we can see from the nonzero
transmission in Fig.~\ref{Fig:N4T}(b), because we have only $N=4$
FG pairs.  Thus our result shows that the condition of occurrence
of the of AP is less stringent than we would have expected
originally.~\cite{Niu90}  In other words, the pumping effect of
our FAG configuration is robust.

It is also worth pointing out that the pumped currents are
positive in Fig.~\ref{Fig:N4C}, showing that the net number flux
of the pumped electrons is from right to left. This is consistent
with the propagation direction of the electrostatic wave in
$V(x,t)$.~\cite{Tang01}

\subsubsection{tuning split-gate}

\begin{figure}[TtH]
\includegraphics[width=.42 \textwidth,angle=0]{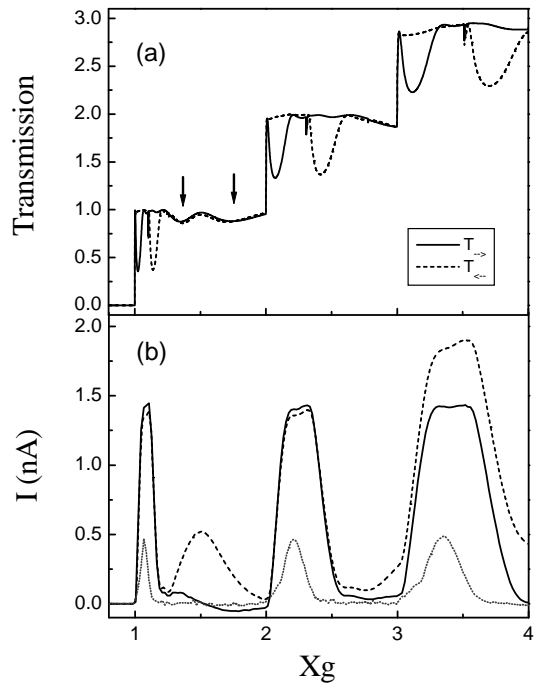}
\caption{The dependence on subband level spacing $\Delta
\varepsilon$ of (a) the total current transmission coefficient,
and (b) the pumped current. The abscissa is depicted by
Eq.~(\ref{eq:xg}) where $\mu = 0.049$ and $N = 4$. Pumping
frequency $\Omega =0.0084$ in all curves except for the dotted
curve in (b), where $\Omega =0.0014$. Parameters $\phi = \pi/2$
and $\alpha = 1/4$ for all curves except for the dashed curve in
(b), where $\alpha=1/5$. In (a), the solid (dashed) curve is for
$T_{\rightarrow}(X_g)$ $\left[T_{\leftarrow}(X_g)\right]$, and
contributions from the second Fourier component of $V(x,t)$ are
indicated by arrows.} \label{Fig:Xg}
\end{figure}

Thus far, we have explored the dependence of the FAG pair's QCP
characteristics on $X_{\mu}$ by the use of the \textit{back-gate
technique}. Another way of tuning the QCP characteristics is via
the modulation of the channel width (or subband level spacing
$\Delta \varepsilon$). This can be realized experimentally by the
use of the so-called \textit{split-gate technique}. Hence we
present, in Fig.~\ref{Fig:Xg}, the transverse confinement
dependence of both the total current transmission coefficients and
the pumped current. The transverse confinement is depicted by \beq
X_{g}=\frac{\mu}{\Delta\varepsilon} +\frac{1}{2}, \label{eq:xg}
\eeq which is linearly related to the effective channel width, and
that its integral value corresponds to the number of propagating
subbands in the channel. In this mode of tuning the QCP
characteristics, $\mu$ is kept fixed.

In Fig.~\ref{Fig:Xg}(a), except for $\mu$, which is fixed at
0.049, and $\omega_{y}$, which varies with $X_{g}$, other
parameters such as $\Omega=0.0084$, $\phi=\pi/2$, and $\alpha=1/4$
are the same as in Fig.~\ref{Fig:N4T}(a). The solid (dashed) curve
is for $T_{\rightarrow}$ ($T_{\leftarrow}$). Both the BS and the
time-dependent Bragg reflection features are found. The expected
locations of the BS, given by the expression \beq
X_{g}=\frac{1}{2}+ \left(n+\frac{1}{2}\right)
\frac{\mu}{\mu-\Omega}, \eeq are at 1.1, 2.3, and 3.5, and they
match the BS locations in Fig.~\ref{Fig:Xg}(a) perfectly. Here $n$
is the subband index. The expected locations of the valleys,
associated with the time-dependent Bragg reflection, are given by
the expression \beq X_{g}=\frac{1}{2}+ \left(n+\frac{1}{2}\right)
\frac{\mu}{\mu-k_{\pm}^{2}}, \eeq thus they should be at $X_g$ =
1.03, 2.1, 3.14 for $T_{\rightarrow}(X_g)$, and at $X_g$ = 1.15,
2.4, 3.73 for $T_{\leftarrow}(X_g)$. Again, they match the valley
locations in Fig.~\ref{Fig:Xg}(a) remarkably.

Besides, there are in Fig.~\ref{Fig:Xg}(a) two additional valley
structures, indicated by arrows, at which $T_{\rightarrow}(X_g)$
and $T_{\leftarrow}(X_g)$ fall one on top of the other. These
structures do not contribute to the pumped current, and they are
due to the time-dependent Bragg reflection from the second order
Fourier component of $V(x,t)$. The second Fourier component of
$V(x,t)$ is in the form of a standing wave, given by
$\cos(2Kx)[\cos\Omega t+\sin \Omega t]$. That both of the
additional valleys all appear in $T_{\rightarrow}(X_g)$ and
$T_{\leftarrow}(X_g)$ can be understood from the fact that more
resonant coupling conditions come into play for the case of
standing wave. The resonant coupling conditions are
$\varepsilon_{k}=\varepsilon_{k\pm 2K}\pm\Omega$, and
$\varepsilon_{k}=\varepsilon_{k\pm 2K}\mp\Omega$. As such, the
valley locations are given by the expression \beq
X_{g}=\frac{1}{2}\left[1+\frac{\mu}{\mu-\epsilon_{\pm}}\right]
\eeq for $n=0$, and for
$\epsilon_{\pm}=[K(1\mp\Omega/(2K)^{2})]^{2}$. Accordingly, these
$2K$ time-dependent Bragg reflection valley locations are expected
to be at 1.36 and 1.73, which coincide with the two additional
valleys in Fig.~\ref{Fig:Xg}(a), which are indicated by arrows.
We note, in passing, that contributions from higher Fourier
components diminishes, as is seen by comparing the valleys from
the first and the second Fourier components of $V(x,t)$.

The $X_{g}$ dependence of the pumped current for the case in
Fig.~\ref{Fig:Xg}(a) is represented by the solid curve in
Fig.~\ref{Fig:Xg}(b). The peaks have flat tops because the valleys
in the corresponding $T_{\rightarrow}(X_{g})$,
$T_{\leftarrow}(X_{g})$ are well separated. The pumped current for
$\Omega=0.0014$, the same frequency as in the case of
Fig.~\ref{Fig:N4T}(b), is depicted by the dotted curve in
Fig.~\ref{Fig:Xg}(b). The peaks are not flat-topped and the
magnitudes are much smaller because the transmission valleys
overlap. For comparison, we also present the case when parameter
values differ slightly from that of the optimal choice. As is
shown by the dashed curve in Fig.~\ref{Fig:Xg}(b), where all
parameters are the same as for the solid curve except that
$\alpha$ is changed from 1/4 to 1/5, the basic pumped current
peaks in the solid curve remain intact. This demonstrates the
robustness of the QCP against the deviation in values of the
configuration parameters from the optimal choice.

Interestingly, there are two additional features in the dashed
curve of Fig.~\ref{Fig:Xg}(b): namely, an additional pumped
current peak at $X_{g}=1.5$, and an increase in the peak value for
the pumped current near $X_{g}=3.5$. That both of these features
are found to arise from the second Fourier component of $V(x,t)$
is supported by the outcome of our analysis performed upon the
Fourier component of $V(x,t)$.  This method of analysis has thus
far been successful in providing us insights on the pumping
characteristics presented in this work. The $m$-TtH Fourier
component of $V(x,t)$, apart from a constant factor, is given by
the form
\begin{eqnarray}
{\cal V}_{m}&=&\left\{[\cos (m\pi\alpha)-{\rm
 sin}(m\pi\alpha)]\,\cos [mKx'-\Omega t-\pi/4]\right. \nonumber \\
&+&\left.  [\cos (m\pi\alpha)+ \sin (m\pi\alpha)]\,\cos
[mKx'+\Omega t+\pi/4]\right\},
 \nonumber \\ & &
\end{eqnarray}
where $x'=x-\delta x/2$. ${\cal V}_{m}$ consists, in general, of
waves propagating in both left and right directions. But when
$\alpha=1/4$, as we have discussed before, ${\cal V}_{1}$ becomes
a pure left-going wave and ${\cal V}_{2}$ becomes a pure standing
wave. The case of $\alpha=1/5$, however, have both ${\cal V}_{1}$
and ${\cal V}_{2}$ consisting of waves in opposite propagation
directions. Therefore, in contrast with the $\alpha=1/4$ result,
additional contributions from the $2K$ Bragg reflection are
expected for the case $\alpha=1/5$. This additional contribution
should peak at the mid-point between two transmission valleys for
the $2K$ Bragg reflections, and the expression for $X_{g}$ is
given by \beq X_{g}=\frac{1}{2}+\left(n+\frac{1}{2}\right)
\frac{\mu}{\mu-\epsilon_{M}}, \eeq where
$\epsilon_{M}=K^{2}+(\Omega/2K)^{2}$. For the case of the dashed
curve in Fig.~\ref{Fig:Xg}(b), the values of $X_{g}$ = $1.54$ and
$3.6$ are shown to match the locations of the additional features
nicely. Finally, we can extract information of the sensitivity of
the pumped current characteristics to $\alpha$ by looking at the
coefficients of the left-going and right-going waves in ${\cal
V}_{m}$. For $\alpha=1/5$, the coefficients of ${\cal V}_{1}$ for,
respectively, the right-going and the left-going waves are $0.22$
and $1.4$. This shows that ${\cal V}_{1}$ is still dominated by
the left-going wave and thus explains the tiny modifications to
the pumped current peaks at $X_{g}$ = $1.1$, and $2.3$. But for
${\cal V}_{2}$, the coefficients for, respectively, the
right-going and the left-going waves are $-0.95$ and $1.57$. This
shows that ${\cal V}_{2}$ deviates quite significantly from that
of a standing wave, and so explains that the additional peaks from
the $2K$ Bragg reflections are quite large.

\section{DISCUSSION AND SUMMARY}

It is interesting to note in passing that our proposal of the FGA
pair configuration is different, in three aspects, from the
voltage lead pattern proposed earlier by Niu.~\cite{Niu90} First
of all, the pumping mechanisms to which the configurations are
catering to are different. It is the mechanism of translating the
Wannier functions in a given Bloch band in
Ref.~\onlinecite{Niu90}, while it is the mechanism of the
time-dependent Bragg reflection in this work. The former mechanism
is adiabatic by nature but the latter mechanism is shown, in this
work, to hold in both the adiabatic and non-adiabatic regimes.

Second, the configurations are different in the number of sets of
voltage leads invoked. A third set of voltage leads was instituted
by Niu to fix the Fermi energy at the middle of the {\it
instantaneous energy gap\/} in order to maintain the adiabaticity
of the pumping. Since our interest here is on the general pumping
characteristics, including, in particular, their dependence on the
Fermi energy, it suffices us to consider a simpler
configuration---the FGA pair configuration. Third, the number of
voltage lead expected, and needed, in a voltage lead set is
different.  Our results demonstrate the resonant nature of the
time-dependent Bragg reflection, and that the pumping
characteristic is robust---requiring only a FGA pair with small
$N$.  Hence the FGA pair configuration proposed in this work
should be more accessible experimentally.

In conclusion, we have proposed a finger-gate array pair
configuration for the generation of quantum charge pumping. Detail
pumping characteristics have been analyzed, the robustness of the
time-dependent Bragg reflection in QCP has been demonstrated, and
the pumping mechanism is understood.

\begin{acknowledgements}
This work was funded by the National Science Council of Taiwan
under Grant Nos. NSC91-2112-M-009-044 (CSC) and
NSC91-2119-M-007-004 (NCTS).
\end{acknowledgements}

\end{document}